# Signal Attenuation through Foliage Estimator (SAFE)

Mathieu Châteauvert, Jonathan Ethier, Pierre Bouchard
Communications Research Centre Canada (CRC)
Ottawa, Ontario, Canada
mathieu.chateauvert, jonathan.ethier, pierre.bouchard @ ised-isde.gc.ca

*Abstract*— The SAFE tool is an open-source Radio Frequency (RF) propagation model designed for path loss predictions in foliage-dominant environments. It utilizes the ITU-R P.1812-6 model as its backbone, enhances predictions with the physics-based Radiative Energy Transfer (RET) model and makes use of high-resolution terrain and clutter elevation datasets.

*Keywords—propagation, path loss modelling, elevation data, foliage loss modelling, rural communications, hybrid models, open source*

## I. Introduction

The Signal Attenuation through Foliage Estimator or SAFE tool is an open-source propagation model for RF path loss predictions in foliage-dominant environments. It has been observed [1] that various open-source propagation models may encounter challenges in providing predictions with foliage along the signal path. The SAFE tool aims at helping any individual, business, or academic access modern propagation models, targeting foliage-dominated areas. SAFE uses ITU-R P.1812-6 [2] as the backbone of the model, as it is an internationally recognized and recommended model and the source code is openly accessible. While P.1812 serves as one of the most proficient model for path loss prediction in foliage-dominant areas, providing reasonable performance in terms of root mean squared error (RMSE) in the range of 13-16 dB, SAFE yields an approximate 3 dB improvement in the RMSE. SAFE is enhanced with a physics-based foliage model known as the RET model [1]. With this addition, the SAFE tool is particularly well-suited for rural environments due to its reliance on a foliage attenuation model, with limited use in urban settings due to its incompatibility with buildings. This paper provides an overview of the SAFE tool, its implementation, and the key components that make it a reliable propagation model in foliage-dominant environments. The source code for the SAFE tool was made freely available [3] and a user's manual is provided at the root of the repository.

## II. Propagation Model Components

The SAFE tool relies on ITU-R P.1812-6 and the RET model to predict path loss, drawing essential environmental information from the High-Resolution Digital Elevation Model (HRDEM) dataset.

### A. High-Resolution Digital Elevation Model

The High-Resolution Digital Elevation Model [4] is a height elevation dataset for Canada. HRDEM is a highly detailed map having a resolution of 1m. HRDEM is derived from airborne Light Detection and Ranging (LIDAR) data from 2015 to the present. It includes two components, namely, the Digital Terrain model (DTM) and the Digital Surface Model (DSM). The DTM represents the height above sea level of the land, including mountains, and roads, while the Digital Surface Model (DSM) represents the height above sea level of objects above the terrain, such as vegetation, buildings, and other types of clutter. The difference between DSM and DTM results in the height of clutter above the terrain.

While this paper is presenting a solution using high-resolution clutter data from Canada, SAFE can leverage the highest-resolution terrain and clutter data where available, while also resorting to lower-resolution data when higher resolution is not accessible. In cases where there is insufficient coverage from HRDEM, the SAFE tool seamlessly makes use of the lower-resolution Canadian Digital Elevation Model (CDEM) [5] dataset. This information (primarily HRDEM and CDEM as a fallback option) is used as inputs to the P.1812 propagation model. However, the RET model requires and only improves SAFE's prediction when high-resolution elevation data is used, as it does not effectively utilize lower-resolution data.

### B. ITU-R P.1812-6 Propagation Model

International Telecommunication Union - Radiocommunication Sector (ITU-R) Recommendation P.1812-6 [2] is a propagation prediction model suitable for terrestrial point-to-point or point-to-area services in the frequency range of 30 to 6000 MHz. The model is appropriate for links having path lengths ranging from 0.25 km to 3000 km, with both Transmitter (Tx) and Receiver (Rx) situated within 3 km above ground. Unlike other general-purpose propagation models such as the Longley-Rice model [6], P.1812 makes use of the clutter information above ground (i.e., buildings and foliage). In SAFE, the time percentage '$p$' and location percentage '$p_L$' are both set at 50%. P.1812 is used with both terrain elevation and clutter data derived from HRDEM. Path profiles between a transmitter and a receiver are described using three sets of data of equal size. For each profile point, inputs include (1) the distance from the transmitter, (2) the terrain height above sea level and (3) the representative clutter height. As per ITU-R Recommendation, SAFE uses equally-spaced profile points with a fixed resolution of 30 m, which after experimentation was confirmed to be an optimal resolution. For each profile point, clutter is detected based on the difference in height between DSM and DTM data. For all paths, the first and last element of the representative clutter height vector must be 0 so as to prevent the P.1812 model from overestimating diffraction loss at the endpoints. Table 1 presents the different clutter types of P.1812 and the representative clutter heights for each class. P.1812 applies a threshold to the clutter such that all clutter heights below said threshold is set to zero and all clutter heights equal to and above said threshold is set equal to the representative clutter height for the associated clutter type. Note that P.1812-6 does not account for propagation through foliage; this task is handled by the RET model, described next.

TABLE I. DEFAULT REPRESENTATIVE CLUTTER HEIGHTS FOR EACH CLUTTER TYPE (P.1812)

| Clutter Type | Representative Clutter Height (m) |
|---|---|
| Water/Open/Rural | 0 |
| Suburban | 10 |
| Urban/Trees/Forest | 15 |
| Dense Urban | 20 |

## C. RET Model for Estimating Propagation Loss through Foliage Clutter

To estimate the attenuation *through* a given expanse of foliage, SAFE determines whether the direct path between the Tx and Rx is obstructed by clutter. The loss through foliage depends on the tree species, the total foliage depth, the incidence angle, the beamwidth of the receiving antenna and the frequency of operation. SAFE currently assumes that all clutter is foliage, and uses a default for a given tree species and Rx beamwidth to determine the attenuation due to foliage in decibels (dB). SAFE relies on a model based on the theory of Radiative Energy Transfer [1,7]; this RET model for slant paths yields reasonably accurate predictions of loss through foliage in the microwave and millimetre wave bands. The grazing incidence version of the RET model was adopted by the ITU-R [1].

The predicted loss through foliage versus distance at various angles of incidence Θ is shown in Figure 1 for American Plane trees in leaf at 3.5 GHz (see Tables 5-8 of [1] for the input parameters). The foliage loss shows a steep initial attenuation rate (dB/m) curve over the first few meters of propagation through foliage and a lower attenuation rate at larger distances as the dominant propagation mode changes from the strongly attenuated direct-path mode to a multiple-scatter mode that is much less attenuated [7]. Note that the angle Θ refers to the angle at which the signal encounters the top of the foliage, is measured with respect to the local horizontal. A grazing incidence (Θ = 0) means that the signal enters the foliage parallel to the ground. Foliage depths over 25 m results in a predicted loss greater than 30 dB for grazing incidence of 30° (red curve), and is even larger at 60° (blue curve). Due to this excessive loss, the through-signal component is unlikely to be the dominant one for deep regions of foliage such as the ones that can be encountered in rural regions.

Other propagation paths around and underneath vegetation need to be considered: the former are diffracted over and around trees and the latter consists of ground reflections. Moreover, since these forests are typically much wider than they are tall and that ground reflections are likely negligible, the top diffracted path is probably dominant over the side-diffracted ones.

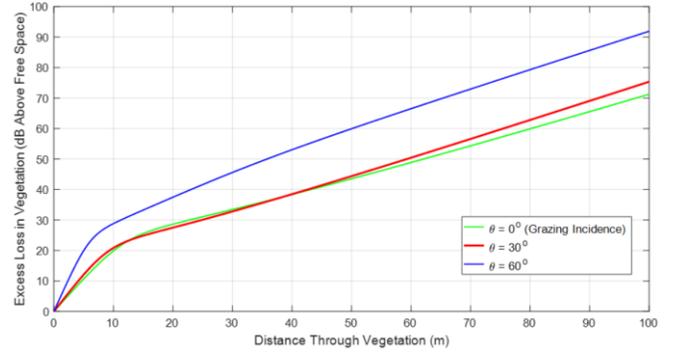

Fig. 1. Predicted loss through American Plane Trees in leaf at 3.5 GHz based on the RET model

## III. SAFE TOOL ARCHITECTURE

The architecture of the SAFE tool is shown in Figure 2. The SAFE tool generates a 2D elevation profile from the HRDEM [4] dataset between the transmitter and receiver (see Figure 3), which is then fed into two models: the P.1812 path loss model and the RET model. These models compute two propagation loss values (one for propagation losses over terrain, the other representing propagation losses through foliage), which are then combined to generate the SAFE path loss prediction.

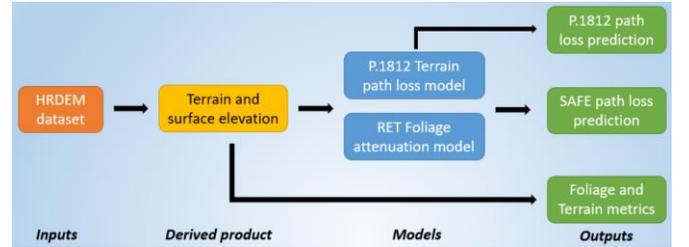

Fig. 2. SAFE Tool Architecture

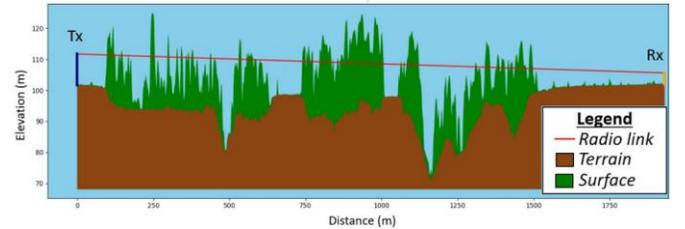

Fig. 3. Representative 2D Path Profile for a Link Extracted from HRDEM data

We operate the P.1812 model in two modes: (1) the ITU-R recommended approach where the clutter is included in the calculation of path loss (forming the baseline to compare against) and (2) with no clutter in the path loss calculation since all of the clutter losses are assumed to be due to foliage, which is alternatively computed by the RET model. The SAFE Path Loss is computed using equation 1:

$$PL_{SAFE} = PL_{P.1812(NoClutter)} + min(RET_{loss}, RET_{Limit}) \quad (1)$$

The value $RET_{loss}$ corresponds to predictions from the RET model and assumes the input parameters of an American Plane

(in leaf) tree at 3.5 GHz [1], and the depth of foliage is computed as the total depth through the clutter intercepting the direct ray path, as determined via the HRDEM clutter information. The RET parameters for the American Plane are used nominally as a deciduous-type tree, which is the most common family of trees in the areas the SAFE tool will be validated against. Future work will extract the tree species of the clutter and apply the correct RET model parameters for a variety of different tree types, thereby improving loss estimates.

Note that we limit the path loss contribution due to foliage clutter by the limit $RET_{limit}$, the value of which is determined from validation studies, as discussed in the next sections.

## IV. VALIDATION OF MODEL IN RURAL, SPARSELY FORESTED ENVIRONMENT

### A. Rural, Sparsely Forested Measurement Data

In November 2013, a series of drive test measurements were performed by the Communications Research Center (CRC) in rural Kanata and Carp, Ontario to analyze the mid-band frequencies (725, 2669 & 4909 MHz) characteristics. The data collection spanned over two days, during which various routes were traversed collecting over 70k individual measurements. The area, right outside the boundaries of Ottawa, Ontario, encompassed agriculture fields and sparse forests. For days 1 and 2, the antenna height used was respectively 16m and 6m. The receiver antenna height was 2.5 m for all measurements.

### B. RET Model Loss Limit

As the total foliage depth increases, the RET model predicts very large losses as seen in Figure 1, which can limit its usefulness. The algorithm for assessing the top diffracted component is still under development. In the interim, following experimentations, a limit of 30 dB was imposed on heavily forested rural areas, and a 20 dB limit was imposed on semi-rural areas. Fig. 4 presents the root mean square error (RMSE) for both region types for a range of RET model dB limits. Despite the minimum error occurring when the limit is set at 12dB, the model employs a 20dB limit to avoid setting it too low and making it too data-specific.

### C. Tree-Growth Factor

Merging datasets from different time periods can create challenges. The measurement data were obtained seven years (2013) prior to the gathering of the elevation data (2020), indicating that there may have been changes in the environment during that time period. Although the terrain remained mostly unchanged, trees are expected to have grown to some extent. To address this, a moderate Tree-Growth Factor of *0.5m/year* [8] was subtracted from the elevation surface to account for the potential increase in tree height, as we assume most of the surface data were foliage in the area.

### D. Rural, Sparsely Forested Validation Results

Table II presents the RMSE between the measured path loss values and the model's predictions; these results show that the SAFE tool outperforms the P.1812 model. The use of P.1812 with suburban clutter results in the lowest RMSE for this model. The addition of the Tree-Growth Factor with SAFE yields improvements in the RMSE, reducing it by 1.2 dB. This improvement provides valuable insights into the potential timestamps disparities that may arise between measurements and elevation date-of-capture.

Unsurprisingly, *P.1812 – No Clutter* results in a general underprediction of the path loss, as can be seen in *Fig.5*. Using P.1812 along with representative clutter data or including the RET model results in a positive mean error, shifting from underpredicting the path loss to overpredicting it. Generally, an overpredicting path loss model is preferable as it accounts for uncertainties and variations in the real-world environment, providing a worst-case prediction. If the actual signal strength is higher than predicted, it ensures better coverage than expected.

TABLE II. RMSE FOR DIFFERENT MODELS IN RURAL, SPARSELY FORESTED MEASUREMENT DATASET

| Model | RMSE (dB) |
|---|---|
| P.1812 – No Clutter | 14.6 |
| P.1812 – Suburban Clutter Classification | 13.4 |
| P.1812 – Urban Clutter Classification | 17.2 |
| SAFE tool | 11.7 |
| SAFE tool (Tree-Growth Factor) | 10.5 |

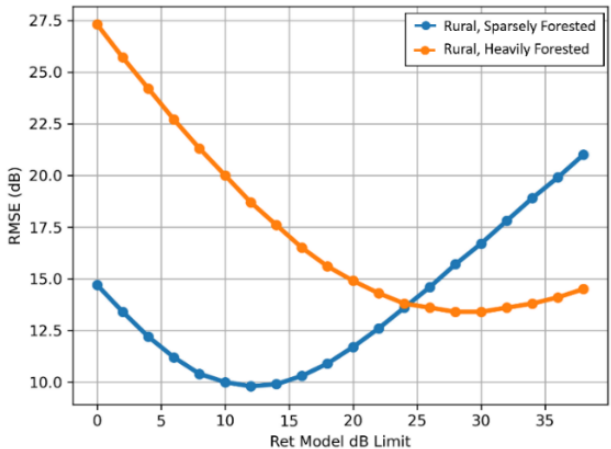

Fig. 4. Impact of Changing the RET model dB limit

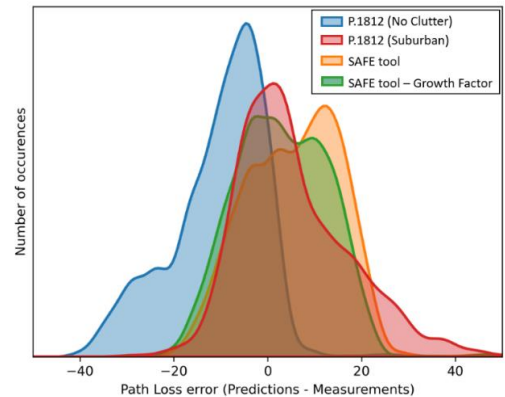

Fig. 5. Prediction error histogram for Rural, Sparsely Forested measurements

Since the data were collected during drive tests and to evaluate the performance of each model, the validation process [9] consisted of grouping data into small bins and comparing the median measured path loss ($PL_M$) with the median predicted path loss ($PL_P$). The data were grouped into small bins using geohash8-sized regions (38.2 x 19.1 meters). For each bin, all values were utilized to calculate the median $PL_M$ and $PL_P$.

To ensure the reliability of the results, a bin was considered valid if it met two conditions : (1) contained at least three measured path losses; (2) the median path loss was lower than or equal to the maximum path loss minus a 6 dB margin (maximum path loss as derived from the Tx Effective Isotropic Radiated Power (EIRP) minus the noise floor of the measured power). All valid bins were used to compute the RMSE using the standard equation 2 :

$$RMSE = \sqrt{\sum_{i=1}^{n} \frac{(PL_M - PL_P)^2}{n}} \quad (2)$$

## V. Model Validation in Heavily Forested Environment

### A. Rural, Heavily Forested Measurement Data

A proprietary set of measurements was used for the validation of the SAFE tool. The precise locations and numbers of the measurements remain obfuscated, but they represent a very large collection of RF measurements at mid-band frequencies and in heavily foliage-dominant regions of Canada.

### B. Rural, Heavily Forested Validation Results

The SAFE tool obtains an RMSE of 13.4 dB on the rural, heavily forested measurement dataset, with a prediction error histogram shown in Fig 6. Additional comparison details for the various modes of P.1812 are shown in Table III. Note that, unsurprisingly, the P.1812 model with no clutter underpredicts the path loss by a very large margin on average due to the abundance of foliage clutter present in some rural regions of Canada. Note also that we do not include predictions with tree-growth corrections, since the measurement and elevation data are aligned in time. The precise count of the measurements is hidden for non-disclosure reasons, but suffice it to say that the number of measurements represents ample validation of the tool.

TABLE III. RMSE FOR THE DIFFERENT MODELS IN THE RURAL, HEAVILY FORESTED MEASUREMENT DATASET

| Model | RMSE (dB) |
|---|---|
| P.1812 – No Clutter | 27.3 |
| P.1812 – Suburban Clutter Classification | 16.1 |
| P.1812 – Urban Clutter Classification | 16.4 |
| SAFE tool | 13.4 |

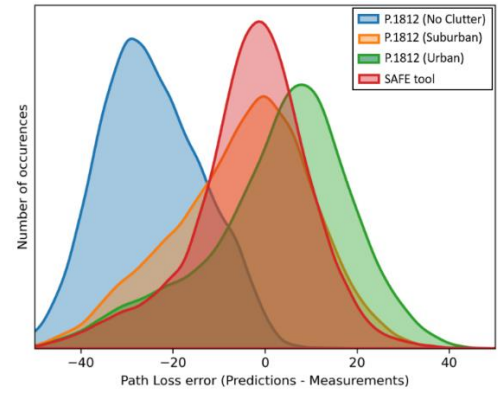

Fig. 6. Prediction error histogram for rural, Heavily Forested measurements

## VI. Concluding Remarks

The SAFE tool combines a well-established path loss model, P.1812, with a physics-based model of foliage loss, the RET model. Working together, these models were shown to be effective in predicting path loss in rural regions, in both sparsely and densely forested areas, with respective RMSE of 10.5 dB and 13.4 dB. It was further shown that the SAFE tool approach of modelling clutter loss in foliage-dominated environments using the RET model offers improvements over using P.1812 alone. Future work will incorporate the ability to select different tree species for the RET model, as well as accounting for the diffracted component associated with foliage-based clutter loss as a complement to the already calculated through component.

### Acknowledgment

The authors would like to thank our CRC colleagues Guillaume Couillard, Amir Basri, Phil Vigneron, Amirah Coja and Mandy Kemp for their help in testing and releasing the SAFE tool.

### References

[1] P.833: Attenuation in vegetation, 2023-07-18, [Online] https://www.itu.int/rec/R-REC-P.833-10-202109-I/en

[2] P.1812: A path-specific propagation prediction method for point-to-area terrestrial services in the frequency range 30 MHz to 6 000 MHz, 2023-07-18, [Online] https://www.itu.int/rec/R-REC-P.1812-6-202109-I/en

[3] CRC. "SAFE Tool" 2023. [Online] https://github.com/ic-crc/SAFE-Tool

[4] High Resolution Digital Elevation Model (HRDEM), NRCan, 2023-07-18, [Online] https://open.canada.ca/data/en/dataset/957782bf-847c-4644-a757-e383c0057995

[5] Canadian Digital Elevation Model (CDEM), NRCan, 2023-07-18. [Online]: https://open.canada.ca/data/en/dataset/7f245e4d-76c2-4caa-951a-45d1d2051333

[6] Longley-Rice: A. G. Longley and P. L. Rice, "Prediction of Tropospheric radio transmission over irregular terrain, A Computer method-1968." ESSA Tech. Rep. ERL 79-ITS 67, U.S. Government Printing Office, Washington, DC, July 1968.

[7] R. A. Johnson and F. Schwering, "A transport theory of millimeter-wave propagation in woods and forests," U.S. Army CECOM, Fort Monmouth, NJ, R&D Tech. Rep. CECOM-TR-85-1, Feb. 1985.

[8] M. Dirr, "Manual of woody landscape plants: their identification, ornamental characteristics, culture, propagation and uses" (August 2009), Stipes Pub, LLC.

[9] ITU-R Document 3K/236, *UK Mobile measurement data Preparation and Benchmarking Method*, 2018-06-13, [Online] https://www.itu.int/md/R15-WP3K-C-0236/en